\documentclass[prd,superscriptaddress,nofootinbib,twocolumn]{revtex4-1}
\usepackage{graphicx,natbib}
\usepackage{url}
\usepackage{color}
\usepackage{rotating}
\usepackage{hyperref}
\usepackage{amssymb,amsmath}
\hypersetup{
    colorlinks=true,
    linkcolor=red,
    citecolor=blue,
}

\newcommand{\mnras}{{Mon.~Not.~R.~Astron.~Soc.}}
\newcommand{\aap}{{A\&A}}

\newcommand{\araa}{{Ann.~Rev.~Astron.~Astrophys. }}
\newcommand{\jcap}{{J. Cosmol. Astropart. Phys.}}

\def\gsim{\;\rlap{\lower 2.5pt
 \hbox{$\sim$}}\raise 1.5pt\hbox{$>$}\;}
\def\lsim{\;\rlap{\lower 2.5pt
 \hbox{$\sim$}}\raise 1.5pt\hbox{$<$}\;}

\begin{document}

\title{Geometrical constraints on curvature from galaxy-lensing cross-correlations}
\author{Yufei Zhang}
\email{zyfeee@mail.ustc.edu.cn}
\affiliation{CAS Key Laboratory for Research in Galaxies and Cosmology, Department of Astronomy, University of Science and Technology of China, Hefei, Anhui, 230026, People’s Republic of China}
\affiliation{School of Astronomy and Space Science, University of Science and Technology of China, Hefei, Anhui, 230026, People’s Republic of China}
\author{Wenjuan Fang}
\email{wjfang@ustc.edu.cn. Corresponding author.}
\affiliation{CAS Key Laboratory for Research in Galaxies and Cosmology, Department of Astronomy, University of Science and Technology of China, Hefei, Anhui, 230026, People’s Republic of China}
\affiliation{School of Astronomy and Space Science, University of Science and Technology of China, Hefei, Anhui, 230026, People’s Republic of China}

\begin{abstract}
Accurate constraints on curvature provide a powerful probe of inflation. However, curvature constraints based on specific assumptions of dark energy may lead to unreliable conclusions when used to test inflation models. To avoid this, it is important to obtain constraints that are independent on assumptions for dark energy. In this paper, we investigate such constraints on curvature from the geometrical probe constructed from galaxy-lensing cross-correlations. We study comprehensively the cross-correlations of galaxy with magnification, measured from type Ia supernovae's brightnesses (``$g\kappa^{\rm SN}$''), with shear (``$g\kappa^{\rm g}$''), and with CMB lensing (``$g\kappa^{\rm CMB}$''). We find for the LSST and Stage IV CMB surveys, ``$g\kappa^{\rm SN}$'' , ``$g\kappa^{\rm g}$''  and ``$g\kappa^{\rm CMB}$'' can be detected with signal-to-noise ratio $S/N=104,\ 2291,\ 1842$ respectively.  When combined with supernovae Hubble diagram (``SN'') to constrain curvature, we find galaxy-lensing cross-correlation becomes increasingly important with more degrees of freedom allowed in dark energy. Without any priors, we obtain error on $\Omega_K$ of $0.723$ from ``SN + $g\kappa^{\rm SN}$'', $0.0417$ from ``SN + $g\kappa^{\rm g}$'', and $0.04$ from ``SN + $g\kappa^{\rm g}$ + $g\kappa^{\rm CMB}$'' for the LSST and Stage IV CMB surveys. The last one is more competitive than a Stage IV BAO survey (``BAO''). When galaxy-lensing cross-correlations are added to the combined probe of ``SN + BAO + CMB'', where ``CMB'' stands for Planck measurement for the CMB acoustic scale, we obtain constraint on $\Omega_K$ of $0.0013$, which is a factor of 7 improvement from ``SN + BAO + CMB''. We study improvements in these results from increasing the high redshift extension of supernovae.    
\end{abstract}

\maketitle

\section{Introduction}

The Universe's curvature is one of its fundamental properties, and deserves to be accurately measured. More importantly, stringent constraints on curvature also provide powerful probes of early Universe physics such as inflation \cite{Guth81,Linde82,AlbrechtSteinhardt82}. Though predictions of the inflationary scenario have been shown to be consistent with accurate measurements of the cosmic microwave background (CMB) anisotropies \cite{Planck18-Inflation}, details of the scenario remain to be uncovered by further measurement results including measurement of the curvature. 

In inflation models with a large number of e-foldings, curvature is predicted to be undetectable, i.e., the magnitude of $\Omega_K$---the curvature density parameter---is below $10^{-5}$, the measurement limit from local fluctuations in the spatial curvature within our Hubble volume. While if ordinary slow-roll inflation is preceded by false vacuum decay, potentially observable open curvature can be produced~\cite{Gott82,Bucher+95,Yamamoto+95,Freivogel+06,Bousso+15}. Specifically, the analyses by~\cite{KlebanSchillo12} and \cite{GuthNomura12} find that future detection of closed curvature at the level of $|\Omega_K|\gsim 10^{-4}$ will exclude eternal inflation and pose challenges to the inflationary scenario, while detection of open curvature at the same level will suggest that false vacuum decay happened before the observable inflation. Therefore, accurate measurement of curvature provides a powerful tool to probe inflation.

Current constraints on curvature come mainly from measurements of the Universe's geometry. Due to severe degeneracy between curvature and dark energy, most of these constraints adopt simple assumptions for dark energy. For example, by assuming dark energy to be the cosmological constant, the Planck collaboration obtains $\Omega_K\simeq 0.0007\pm 0.0019$ from their measurements of the CMB temperature, polarization, reconstructed lensing, and external measurements of the baryon acoustic oscillations (BAO) \cite{Planck18}, which is probably the best precision we can achieve today under this assumption. However, since the nature of dark energy remains a mystery \cite{DETF,Frieman+08,CaldwellKamionkowski09,Weinberg+12}, one cannot reach a decisive conclusion when using these constraints to test inflation models. Actually, with the same assumption for dark energy, the Planck team found their measurements for the CMB temperature and polarization alone prefers a closed Universe otherwise \cite{Planck18}, see also \cite{Valentino20,Handley}. While the real reason for this apparent discrepancy remains to be clarified, the situation also stresses the importance of obtaining stringent constraints on curvature that are independent on the uncertainties in our knowledge about dark energy.

In this paper, we investigate the constraints on curvature that are independent on the ``unknown'' properties of dark energy by utilizing geometrical cosmology probes. Specifically, besides the most extensively explored probing techniques such as type Ia supernovae, BAO, and CMB, we study the constraints from the cross-correlation between galaxy (tracers of large-scale structure) distribution and signatures of weak-lensing, for which we comprehensively consider the magnification of the brightnesses of background standard candles such as type Ia supernovae, cosmic shear measured from the distorted images of background galaxies, and the remapping of the CMB fields.

The galaxy-weak lensing cross-correlation itself certainly involves information on structure growth. However, by comparing the signals for the same lensing galaxies but sources at different redshifts, pure geometric information can be extracted \cite{JainTaylor03,HuJain04,Zhang++05}. With more and more ambitious cosmological surveys starting or to start operation, especially the planned stage IV dark energy experiments such as the LSST \cite{LSST}, CSST \cite{CSST,Gong+19}, the stage IV CMB experiment \cite{CMBS4}, this geometrical probe is drawing more attention these days. For example, ratios of distances have recently been measured from real surveys with high significance using the cross-correlation between galaxy and cosmic shear (the so-called galaxy-galaxy lensing), and the cross-correlation between galaxy and CMB lensing \cite{Miyatake+2017,Prat2019}. At the same time, the cross-correlation between galaxy and supernovae magnification itself has not been conclusively detected with current data yet. For recent trials, see \cite{Smith+14,Mitra2018,Sakakibara+2019}. The opportunities with future surveys such as the LSST should be wide-ranging, see \cite{Scovacricchi+2017} and our investigations in Sec.~\ref{sec:lsst} below.

\begin{figure}
\begin{center}
{\includegraphics[angle=0, scale=0.65]{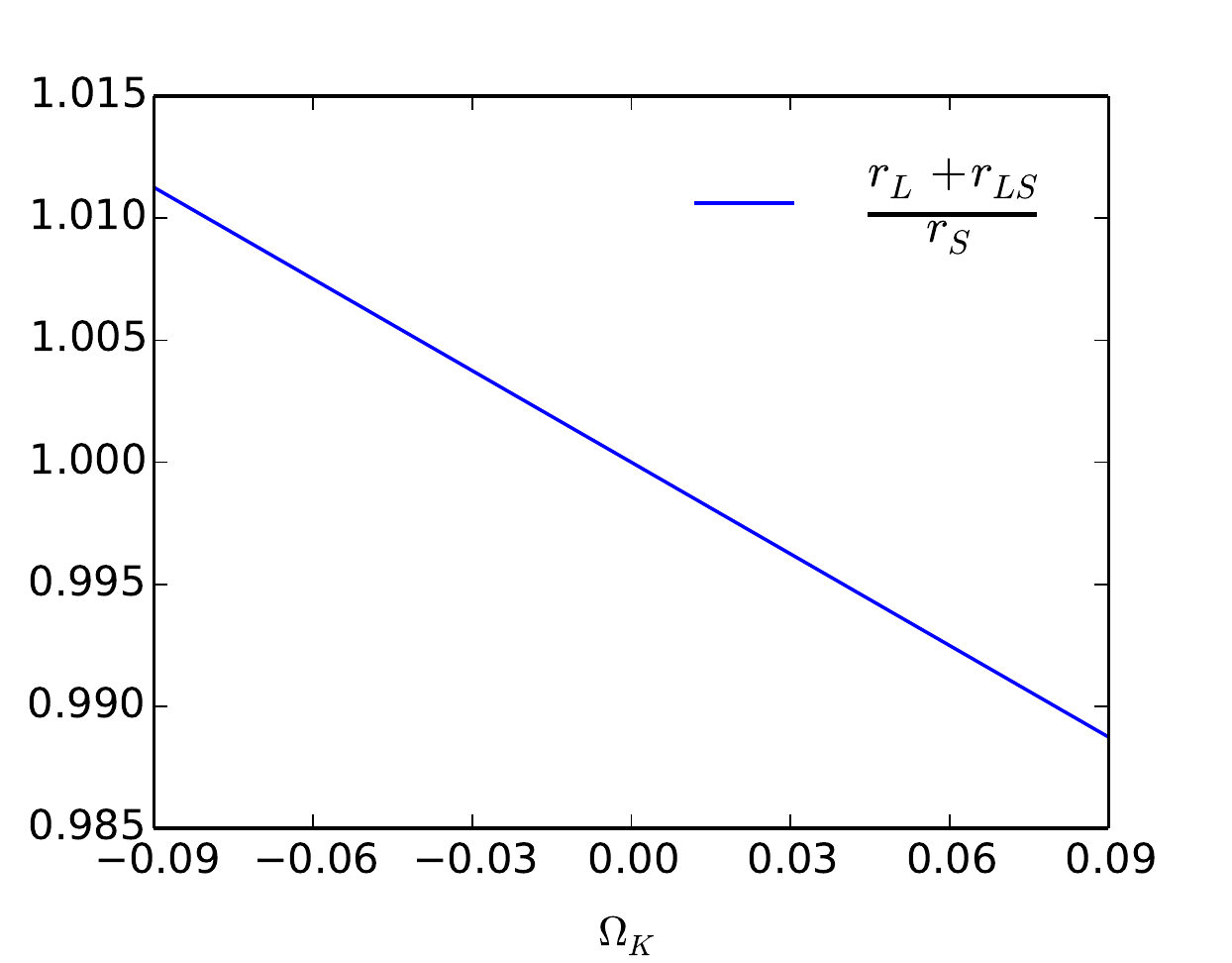}}
 \caption{\label{fig:rLS} The comoving angular diameter distances to the lens $r_L$, to the source $r_S$, and from the lens to the source $r_{LS}$ provide a probe of curvature altogether. Shown in this plot is the dependence on $\Omega_K$ for the ratio of $(r_L+r_{LS})$ to $r_S$, while $r_L$, $r_S$ fixed at $1500$ and $3000 h^{-1}$Mpc respectively. Here, we have used the approximation for $r_{LS}$ when $|\Omega_K|\ll 1$~\cite{Peebles}.}
\end{center}
\end{figure}

Gravitational lensing uniquely probes the angular diameter distance from the lens (rather than from observers at $z=0$) to the source, in addition to the distances to the lens and to the source. It is known that the three distances altogether provide a pure metric probe for curvature~\cite{Bernstein06,Rasanen+14}, see Fig.~\ref{fig:rLS}. In particular, \cite{Bernstein06} proposed to obtain model-independent constraints on curvature by using the geometrical probe constructed from galaxy-galaxy lensing. However, their analysis done in real space adopts an oversimplified assumption that observables along different line of sights are completely independent. In this paper, without directly dealing with the correlations between observables along different line of sights, we perform an analysis in Fourier space that automatically takes into account the correlations. In addition, we extend the analysis to include galaxy-supernovae magnification and galaxy-CMB lensing cross-correlations. We notice that cosmological constraints from the cross-correlation between galaxy and supernovae magnification have been barely explored, if not completely none. We forecast the dark energy independent constraints on curvature from galaxy-weak lensing cross-correlations for Stage IV dark energy experiments, and see whether the desired accuracy level of $10^{-4}$ can be reached when they are combined with other popular geometrical probes.

The rest of this paper is organized as follows. In Sec.~\ref{sec:theory}, we present the formulas we use to forecast the pure geometrical constraints from galaxy-lensing cross-correlations. In Sec.~\ref{sec:results}, we forecast the constraints on curvature with different assumptions for dark energy from the geometrical probe of galaxy-lensing cross-correlations and the ultimate combination with supernovae, BAO and CMB. We discuss our results in Sec.~\ref{sec:discuss} and summarize in Sec.~\ref{sec:summary}.

\section{Theoretical Calculations}
\label{sec:theory}

In this section, we present the theoretical calculations to forecast the pure geometrical constraints from galaxy-lensing cross-correlations. To break the parameter degeneracies, we will combine the constraints with those from the supernovae Hubble diagram, which is available from the same supernovae survey used to measure the galaxy-supernovae cross-correlation, and the calculation of which is given in Sec \ref{sec:SN}. There are great similarity among the three types of galaxy-lensing cross-correlations. The differences are mainly in source redshifts and noises for measuring the lensing signals. Thus, we elaborate the calculations for the galaxy-supernovae cross-correlation, which are much less presented in the literature, while briefly mention those for the other two. We note our calculations in Sec.\ref{sec:gSN} applies to other types of standard candles as well.

\subsection{Type Ia Supernovae}
\label{sec:SN}

The apparent magnitude of a supernova at redshift $z_i$ is given by 
\begin{equation}
m_i=5 \log_{10}\left[H_0 d_L(z_i)\right]+\mathcal{M}+\epsilon_i,\label{eq:mi}
\end{equation}
We introduce $\mathcal{M}_i$ as a quantity involving the supernova's intrinsic luminosity and the Hubble constant $H_0$, and separate $\mathcal{M}_i$ into its mean $\mathcal{M}$ and statistical variation $(\mathcal{M}_i-\mathcal{M})$. The latter is included in $\epsilon_i$, which we use to represent the total variation in $m_i$. The luminosity distance $d_L$ is related to the comoving angular diameter distance $r$ by $d_L=(1+z)r$, while $r$ is given as a function of the comoving radial distance $\chi$ by
\begin{equation}
r(\chi)=\frac{1}{\sqrt{-\Omega_K}H_0}\sin{\left[\sqrt{-\Omega_K}H_0\chi\right]},
\end{equation}
with $\chi$ calculated by $\chi=\int_0^z dz'/H(z')$.

We assume the total number of supernovae discovered from a supernovae survey is $N_{\rm tot}$, and their redshift distribution is $dn/dz$, which we normalize to be 1. To forecast parameter constraints from a supernovae-alone probe --- the supernovae Hubble diagram, following~\cite{LSST}, we neglect all possible correlations between the apparent magnitudes of different supernovae, and for the variance in an individual supernova's apparent magnitude $\sigma_ m^2$, we assume it is totally due to statistical variation of the supernova's intrinsic luminosity, and take it to be a constant. We note that correlations between different supernovae induced by various systematic effects and other types of statistical variations should be taken into account for a more accurate forecast. We then use the Fisher matrix technique to forecast the anticipated constraints, which is constructed as
\begin{equation}
F_{\alpha\beta}=N_{\rm tot}\int dz \frac{dn}{dz}\frac{1}{\sigma_m^2}\frac{\partial\bar{m}(z)}{\partial p_{\alpha}}\frac{\partial\bar{m}(z)}{\partial p_{\beta}}.
\end{equation}
Here, besides the cosmological parameters, our parameter set also includes $\mathcal{M}$ which involves both the mean of the supernovae's intrinsic brightness and $H_0$. In obtaining constraints on the cosmological parameters, we marginalize over $\mathcal{M}$. It can be seen that the parameter constraints will be inversely proportional to $N_{\rm tot}^{1/2}$.

\subsection{Galaxy-lensing cross-correlation}

\subsubsection{Galaxy-supernovae cross-correlation}
\label{sec:gSN}

In the previous section, we have considered the total variation in an individual supernova's brightness at a given redshift comes only from its intrinsic luminosity. In reality, an additional type of variation will be introduced by the magnification effect of gravitational lensing by matter distribution in the foreground~\cite{Frieman96,Wambsganss+97,Holz98,Metcalf99}. Some supernovae are magnified, and some demagnified. In the limit of weak lensing, $\epsilon_i$ in Eq.~(\ref{eq:mi}) will have an additional term of $-5/(\ln{10})\kappa_i$, besides $(\mathcal{M}_i-\mathcal{M})$. Here, $\kappa_i$ is the lensing convergence for a source at the supernova's location. This magnification effect can be statistically measured by cross-correlating the supernovae's brightnesses with the distribution of large-scale structure tracers in their foreground such as galaxies (see e.g., ~\cite{WilliamsSong04,MenardDalal05,Cooray+06}), which we study in this section.

Specifically, we consider the cross-correlation between the following two observables,
\begin{eqnarray}
\delta_n^{2D}(\vec{\theta})&\equiv&\int dz W_L(z)\delta_n(z,\vec{\theta})\label{eq:dn},\\
M(\vec{\theta})&\equiv&\int dz W_S(z)m(z,\vec{\theta}),
\end{eqnarray}
where $\delta_n^{2D}$ and $\delta_n$ are the two and three-dimensional galaxy overdensities; $W_L$ and $W_S$ are the redshift selection functions for the galaxies (``Lens'') and supernovae (``Source'') respectively, both of which have been normalized to be 1, i.e., $\int W(z)dz=1$; 
$M(\vec{\theta})$ represents the average apparent magnitude for the selected supernovae whose angular positions are within a solid angle $d^2\theta$ ($d^2\theta\rightarrow 0$) around $\vec{\theta}$. Intrinsic fluctuations in supernovae's brightnesses drop out in the cross-correlation, and by a Fourier transform, we get the following cross-correlation power spectrum,
\begin{equation}
C_{\ell}^{LS}=-\frac{15}{2\ln(10)}\Omega_m H_0^2\int dz W_L(z)\frac{g_S(z)}{ar^2}P_{gm}(k=\frac{\ell}{r},z),\label{eq:clcross}
\end{equation}
where $\Omega_m$ is the density parameter for matter, $a$ is the scale factor, $P_{gm}$ is the galaxy-matter power spectrum, and $g_S(z)$ is given by,
\begin{equation}
g_S(z)=\int dz'W_S(z')\frac{r(\chi)r(\chi'-\chi)}{r(\chi')}\Theta(\chi'-\chi),
\end{equation}
where $\Theta$ is the Heaviside step function. In our derivation for Eq~(\ref{eq:clcross}), we have used the Limber approximation~\cite{Limber,Kaiser92}.

Galaxy-supernovae cross-correlation potentially provides a pure geometrical probe of the Universe, as is the case for galaxy-galaxy lensing~\cite{JainTaylor03,Zhang++05}. A direct way to see this is by taking the limit that the foreground galaxies are all selected to be at a single redshift, i.e., $W_L(z)\rightarrow\delta(z-z_L)$, then we have $C_{\ell}^{LS}/C_{\ell}^{LS'}\rightarrow g_S(z_L)/g_{S'}(z_L)$, i.e., the ratio of the cross-correlations between these galaxies and supernovae selected with different selection functions, $W_S(z)$ and $W_{S'}(z)$, probes a pure geometrical quantity. Furthermore, in the limit of $W_S(z)\rightarrow\delta(z-z_S)$ and $W_{S'}(z)\rightarrow\delta(z-z_{S'})$, $C_{\ell}^{LS}/C_{\ell}^{LS'}\rightarrow r(\chi_S-\chi_L)r(\chi_{S'})/r(\chi_S)r(\chi_{S'}-\chi_L)$, which directly probes the ratio of the angular diameter distances.

In this section, we investigate the pure geometrical probe constructed from the galaxy-supernovae cross-correlation. We consider a survey of both supernovae and galaxies, and assume the former to be observed to a maximum redshift $z_{\rm max}$, and the latter to a redshift beyond that. For measurements of the cross-correlation, galaxies at $z>z_{\rm max}$ would be of no use, since their cross-correlation with the supernovae would vanish. We divide both the supernovae and galaxies from $z=0$ to $z=z_{\rm max}$ into redshift bins of equal size $\Delta z$. Hereafter, we use ``S'' as the index for the supernova (source) redshift bins, and ``L'' for the galaxy (lens) redshift bins. The measured cross power spectra $C_{\ell}^{LS}$ and $C_{\ell'}^{L'S'}$ have the following covariance,
\begin{equation}
{\rm{Cov}}(C_{\ell}^{LS},C_{\ell'}^{L'S'})=\frac{\delta_{\ell\ell'}^K}{2\ell\Delta\ell f_{\rm sky}}\left(C_{\ell}^{SS'}C_{\ell}^{LL'}+C_{\ell}^{LS'}C_{\ell}^{L'S}\right),\label{eq:cov}
\end{equation}
where $\delta^K$ is the Kronecker delta, $\Delta\ell$ is the size of the multipole bin used to measure $C_{\ell}^{LS}$, and $f_{\rm sky}$ is the fraction of sky covered by the survey. Both the auto power spectra for the supernovae brightnesses $C_{\ell}^{SS'}$ and for the galaxy distribution $C_{\ell}^{LL'}$ have two contributions: one from large-scale structure (``LSS''), and the other from shot noise (``shot''), given by
\begin{eqnarray}
C_{\ell,\rm{shot}}^{SS'}&=&\delta_{SS'}^K\sigma_m^2/\bar{n}_S^{2D},\\
C_{\ell,\rm{LSS}}^{SS'}&=&(\frac{15}{2\ln(10)}\Omega_m H_0^2)^2\int dz\frac{d\chi}{dz} \nonumber\\ &&\times\frac{g_S(z)g_{S'}(z)}{a^2r^2} P_{mm}(k=\frac{\ell}{r},z),\\
C_{\ell,\rm{shot}}^{LL'}&=&\delta_{LL'}^K/\bar{n}_L^{2D},\\
C_{\ell,\rm{LSS}}^{LL'}&=&\int dz W_L(z)W_{L'}(z)(\frac{d\chi}{dz})^{-1}r^{-2}\nonumber\\&&\times P_{gg}(k=\frac{\ell}{r},z),
\end{eqnarray} 
where $\sigma_m^2$, as in Sec.~\ref{sec:SN}, is the variance of an individual supernova's brightness due to its intrinsic luminosity, $\bar{n}^{2D}_S,\bar{n}^{2D}_L$ are the mean angular number densities for supernovae in the ``S''th redshift bin and galaxies in the ``L''th redshift bin respectively, while $P_{mm}$ and $P_{gg}$ are the matter and galaxy power spectra in turn. Note $C_{\ell,\rm{LSS}}^{SS'}$ is the power spectrum of the E-mode shear except for a constant factor of $(5/\ln(10))^2$, see e.g., \cite{Zhang++05, FangHaiman07}. 

Same as before, we use the Fisher matrix technique to forecast parameter constraints from galaxy-supernovae cross-correlation. Since statistical isotropy implies that different multipoles are uncorrelated, the Fisher matrix can be written as a sum of contributions from different multipoles,
\begin{equation}
F_{\alpha\beta}=\sum_{\ell}\sum_{\substack{(LS),\\ (L'S')}}\frac{\partial C_{\ell}^{LS}}{\partial p_{\alpha}}({\rm Cov}_{\ell})^{-1}\frac{\partial C_{\ell}^{L'S'}}{\partial p_{\beta}}\label{eq:gSNfish},
\end{equation}
where $(LS)$ or $(L'S')$ labels distinct cross power spectra. ${\rm Cov}_{\ell}$ is the subblock of the full covariance matrix (Eq.~\ref{eq:cov}) for all the cross power spectra with multipole $\ell$. We note its inverse is proportional to $2\ell\Delta\ell f_{\rm sky}$, hence $F_{\alpha\beta}\propto f_{\rm sky}$, and the parameter constraints will be proportional to $f_{\rm sky}^{-1/2}$.

To extract the pure geometrical constraints, we choose our redshift bins to be narrow enough such that the following approximation (under the limit of $\Delta z \rightarrow 0$) to the cross power spectrum holds to a good accuracy,
\begin{eqnarray}
C_{\ell}^{LS}&\approx& -\frac{15}{2\ln(10)}\Omega_m H_0^2 \frac{r(\chi_S-\chi_L)}{r(\chi_S)r(\chi_L)}\frac{1}{a(z_L)}\nonumber\\&&\times P_{gm}(k=\frac{\ell}{r(\chi_L)},z_L)\Theta(\chi_S-\chi_L),
\end{eqnarray}
where $\chi_S=\chi(z_S)$, $\chi_L=\chi(z_L)$, with $z_S$, $z_L$ representing the mean redshifts of the narrow supernova and galaxy bins respectively. With this approximation, we can easily separate geometrical information ($r(\chi_S-\chi_L)/r(\chi_S)$) from what remains whose prediction typically involves uncertainties in galaxy bias and matter power spectrum in the nonlinear regime, which we hereafter denote as $C_{\ell}^{gm}(z_L)$.
Next, we take the $C^{gm}$s at different multipoles and redshifts also as parameter entries for the Fisher matrix, and marginalize over them for the final geometrical constraints on the cosmological parameters. Including these extra parameters significantly increases the dimension of the Fisher matrix, hence increases the difficulty for its inversion. However, from Eq.~(\ref{eq:gSNfish}), we find that $F_{\alpha\beta}=0$, when $p_{\alpha}$ and $p_{\beta}$ correspond to $C^{gm}$ at different multipoles. This feature greatly simplifies the inversion of the Fisher matrix with the method of ``inversion by partitioning''~\cite{NR1992}.

\subsubsection{Galaxy-galaxy lensing}
\label{sec:gglensing}

Compared to galaxy-supernovae cross-correlation, galaxy-galaxy lensing~\cite{Tyson+84} can be detected with a stronger significance and to a higher redshift, for it is much easier to observe a large number of galaxies to a high redshift than to observe supernovae. In this section, we consider the pure geometrical probe from galaxy-galaxy lensing. 

While galaxy-supernovae cross-correlation is the correlation between the distribution of a foreground galaxy population and magnifications in the background supernovae's brightnesses, galaxy-galaxy lensing is the correlation between the former and distortions in the background galaxies' images caused by weak lensing - the cosmic shear field. To be explicit, galaxy-galaxy lensing is the cross-correlation between $\delta_n^{2D}$, given by Eq.~(\ref{eq:dn}), and $\Gamma^i$, given by the following,
\begin{equation}
\Gamma^i(\vec{\theta})=\int dz W_S(z)\gamma^i(z,\vec{\theta}),
\end{equation} 
where $\gamma^i$ is the shear field estimated through measurements of background galaxies' ellipticities, and ``$i$'' labels the two shear components. Note, $W_S(z)$ here is the selection function for the ``sources'', i.e., background galaxies. Similar to $M(\vec{\theta})$, $\Gamma^i(\vec{\theta})$ represents the average of the shear estimated from background galaxies whose angular positions are within a solid angle $d^2\theta$ ($d^2\theta\rightarrow 0$) around $\vec{\theta}$. 

The shear field can be decomposed into a curl-free E-mode component and a divergence-free B-mode component. With scalar perturbations alone, only the E-mode shear exists, which is equivalent to the lensing convergence. Therefore, we only need to study the cross-correlation between galaxy and the E-mode shear, which we recognize to be exactly the same as the galaxy-supernovae cross-correlation, except for a factor of $5/\ln(10)$. 

In this section, we consider a weak lensing survey. As before, we divide the galaxies into redshift bins of equal size $\Delta z$, and denote the galaxy-galaxy lensing power spectrum as $C_{\ell}^{LS}$. Different from before, ``S'' here labels the redshift bin for the ``source'' galaxies. The expression for the covariance of $C_{\ell}^{LS}$ and $C_{\ell'}^{L'S'}$ remains the same as before, except the following differences: (1) $C_{\ell,\rm LSS}^{SS'}$ has not a factor of $(5/\ln(10))^2$; (2) $C_{\ell,\rm{shot}}^{SS'}$ is now given by
\begin{equation}
C_{\ell,\rm{shot}}^{SS'}=\frac{\delta_{SS'}^K}{\bar{n}_S^{2D}}\int dz W_S(z)\gamma_{\rm rms}^2(z),\label{eq:ggshot},\\
\end{equation}    
where $\bar{n}_S^{2D}$ is the mean angular number density for source galaxies in the ``S''th redshift bin, and $\gamma_{\rm rms}$ is the rms of shear in each component from galaxies' intrinsic ellipticities. In the end, we use the same method as in Sec~\ref{sec:gSN} to forecast the pure geometrical constraints on cosmological parameters from galaxy-galaxy lensing.

\subsubsection{Galaxy-CMB lensing cross-correlation}
\label{sec:gCMBlensing}

The Universe's large-scale structure gravitationally deflects the photons of CMB as well, and thus perturbs the CMB power spectra. The weak lensing convergence for the CMB can be reconstructed from the various lensed CMB power spectra using the minimum variance estimator, which minimizes the reconstruction noise \cite{Okamoto&Hu}. For cosmic shear, the source galaxies are typically distributed across a relative broad range of redshift, and the signals are then averaged over this distribution. However, the CMB photons originate from a very narrow range of comoving distance, thus the source redshift distribution can be approximated as a Dirac $\delta$-function with the value of redshift known to a very precise level. For the cross-correlation between galaxy and CMB lensing, the expressions for the cross power $C_{\ell}^{LS}$ and the covariance between $C_{\ell}^{LS}$ and $C_{\ell'}^{L'S}$ (``S'' here labels the redshift bin for the ``source'' of CMB) remain the same as for the galaxy-galaxy lensing, except $C_{\ell,\rm{shot}}^{SS}$ is now replaced by the CMB lensing reconstruction noise, whose expression is given by, e.g. Eq.(42) in \cite{Okamoto&Hu}.

\section{curvature constraints}
\label{sec:results}

In this section, we present the constraints on curvature from the geometrical probe constructed from galaxy-lensing cross-correlations. In Sec~\ref{sec:lsst}, we forecast the constraints from the galaxy-lensing cross-correlations in combination with the supernovae Hubble diagram. We make our forecast for fiducial surveys mimicking the LSST and Stage IV CMB experiment. In addition to detecting a large number of type Ia supernovae, the LSST will measure both the galaxy distribution and cosmic shear field at the same time. Hence, three of the probes discussed in the last section, i.e., the supernovae Hubble diagram, galaxy-supernovae cross-correlation, and galaxy-galaxy lensing, will be available from the LSST, while the galaxy-CMB lensing cross-correlation can be measured from the overlapping area between the LSST and Stage IV CMB experiment. (We assume the Stage IV CMB survey overlaps completely with the LSST.) In Sec~\ref{sec:BAO}, we add in BAO and CMB to further tighten the constraints. In Sec~\ref{sec:zmax}, we study improvements in the constraints from increasing the high redshift extension of supernovae while keeping their total number fixed.

\subsection{Combination of galaxy-lensing cross-correlation with supernovae Hubble diagram}
\label{sec:lsst}

The supernovae Hubble diagram probes the angular diameter distance from a given redshift to an observer on the earth, i.e., at $z=0$, while the galaxy-lensing cross-correlation can additionally probe the angular diameter distances from a given redshift (the sources' redshift) to observers at all intermediate redshifts with $z\neq 0$ (the lenses' redshifts). Hence, the latter provides important complementary information.

\begin{figure}
\begin{center}
{\includegraphics[angle=0,scale=0.7]{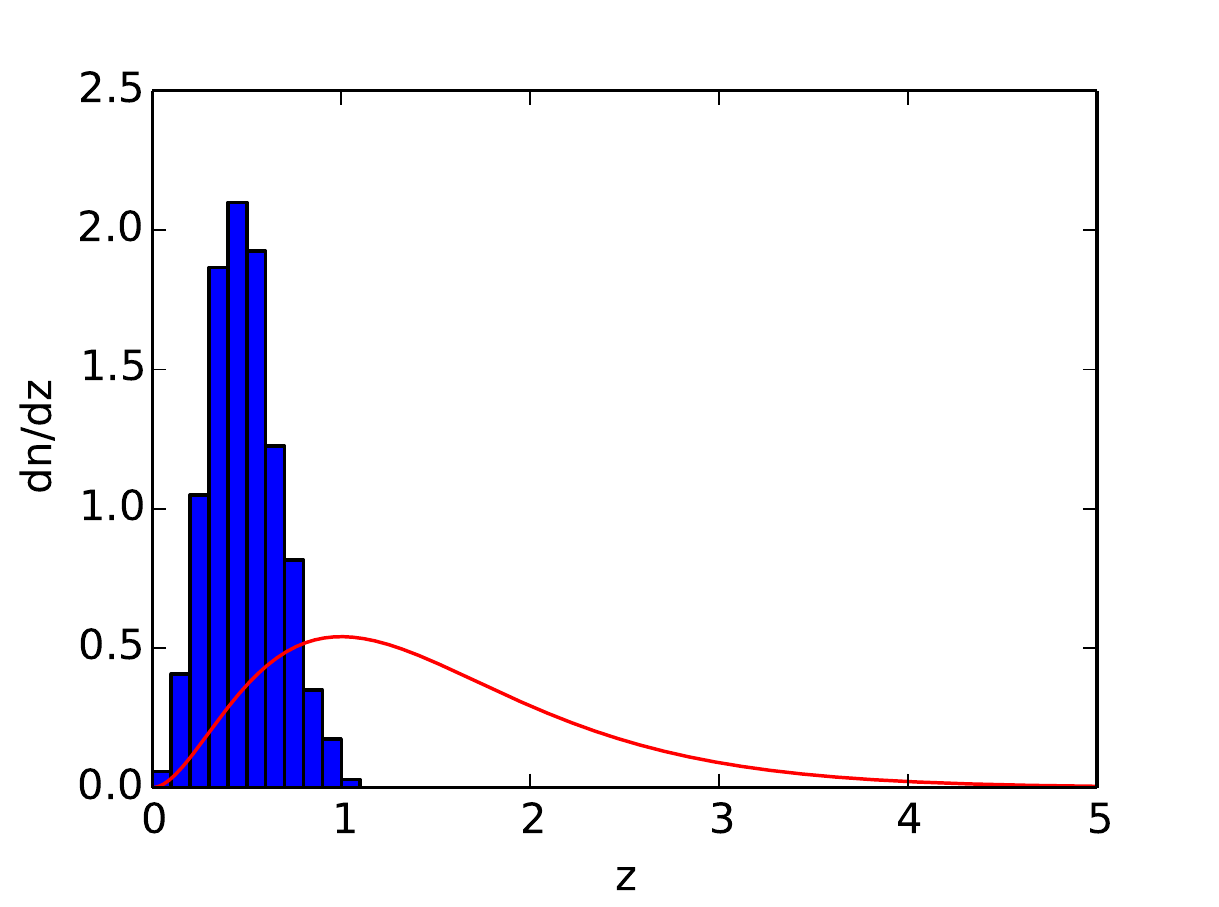}}
 \caption{\label{fig:dndz} Normalized redshift distributions of type Ia supernovae (histogram) and galaxies (solid line) for the LSST.}
\end{center}
\end{figure}

The LSST is about to survey a sky area of approximately 20,000 deg$^2$ for a duration of 10 years starting by 2022 \cite{LSSTSci}. It will detect about half a million type Ia supernovae to a redshift slightly beyond $z=1$, see Fig.~\ref{fig:dndz} for the supernovae's redshift distribution~\cite{LSST}. In the following, we assume the total number of supernovae $N_{\rm tot}$ to be $4\times10^5$~\cite{LSSTSci}, and the rms of their intrinsic brightnesses $\sigma_m$ to be $0.1$~\cite{LSST}. At the same time, the LSST will observe galaxies at an average angular number density of $50$ arcmin$^{-2}$, which is the so-called “gold” sample of the LSST galaxies, the redshift distribution of which is shown as the solid line in Fig.~\ref{fig:dndz}~\cite{LSST}. Among these galaxies, we assume about $60\%$ of them can be used for shear measurement~\cite{LSSTSci}, hence, for galaxy-galaxy lensing, the number density of source galaxies is 30 arcmin$^{-2}$, and we assume $\gamma_{\rm rms}$ has a redshift-independent value of $0.28$~\cite{LSST}. For the Stage IV CMB experiment, we assume a 1 arcmin beam and $1\mu \rm K$ arcmin noise~\cite{CMBS4}, and assume it covers the LSST survey area.

\begin{figure}
\begin{center}
{\includegraphics[angle=0, scale=0.45]{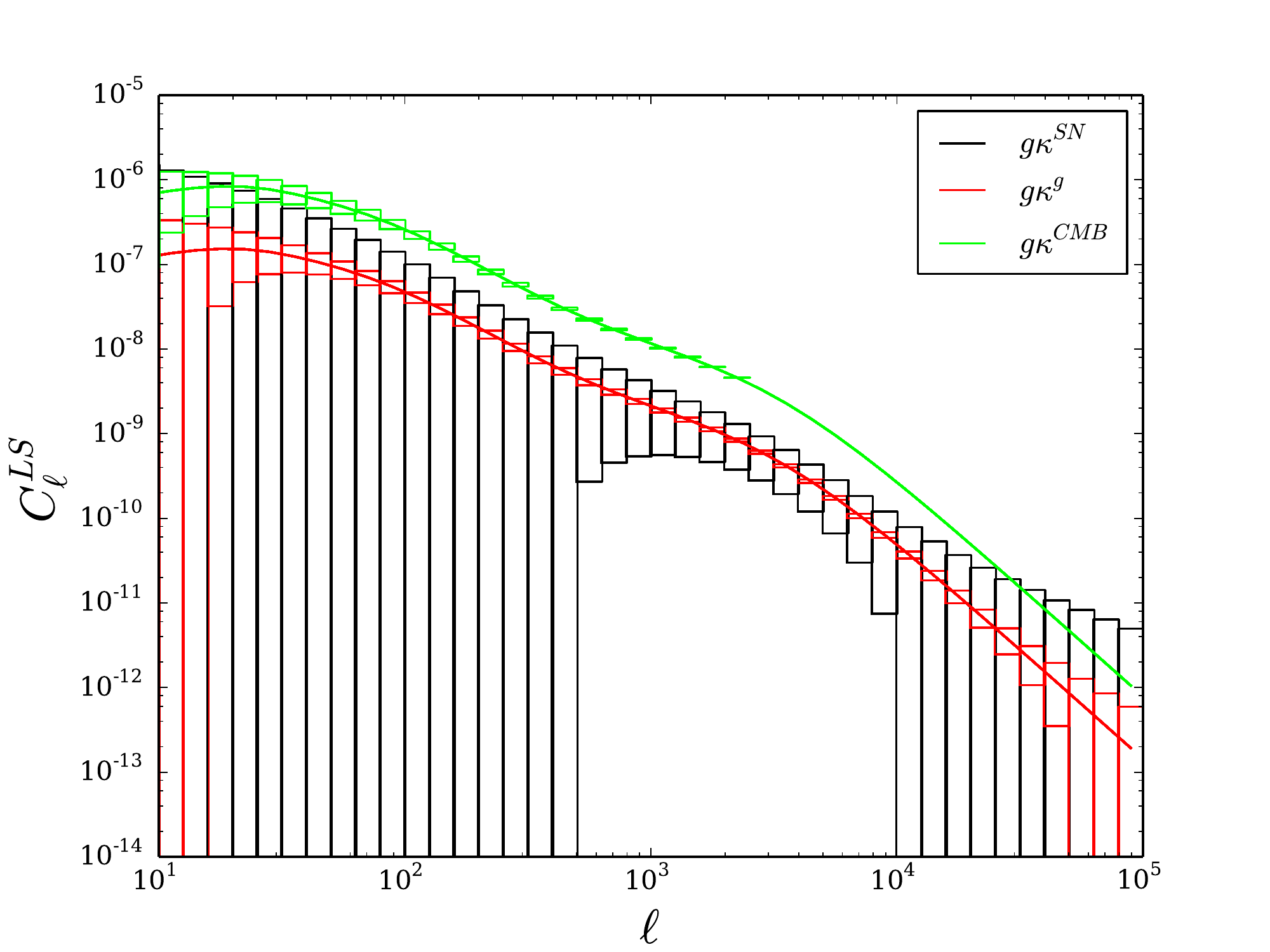}}
 \caption{\label{fig:clLS} The power spectra for the galaxy-supernovae cross-correlation (``$g\kappa^{\rm SN}$'', black), galaxy-galaxy lensing (``$g\kappa^{\rm g}$'', red), and galaxy-CMB lensing cross-correlation (``$g\kappa^{\rm CMB}$'', green). Error bands are forecasted for the LSST and Stage IV CMB experiment. We have divided the galaxy-supernovae cross power and its errors by a factor of $5/\ln(10)$. The lens galaxies are in the redshift bin of z:[0.4, 0.5], while the supernovae and source galaxies are in z:[0.5, 0.6]. The error bands for the galaxy-CMB lensing cross power are plotted up to $\ell=3000$.}
\end{center}
\end{figure}

For the galaxy-lensing cross-correlations, we choose a redshift bin size of $\Delta z =0.1$ for the type Ia supernovae, lens and source galaxies. We note that the recent analysis by \cite{Prat2019} chooses a bin size of $\Delta z =0.15$ for the lens galaxies, and finds the correction from the narrow tracer bin approximation to be negligible compared to their measurement errors. For galaxy-supernovae cross-correlation, we have 11 supernovae bins from $z=0$ to $z_{\rm max}=1.1$ -- the highest supernova redshift, and this gives us 55 ($10\times11/2$) distinct cross power spectra. For galaxy-galaxy lensing, we cut off the galaxy distribution at $z_{\rm max}=4$ in our numerical calculation, which includes $\sim 99\%$ of all the galaxies, and we have $780$ distinct cross power spectra. This also gives us 40 galaxy-CMB lensing cross power spectra. In Fig.~\ref{fig:clLS}, we explicitly show the power spectra for the galaxy-supernovae cross-correlation (``$g\kappa^{\rm SN}$'', black) and galaxy-galaxy lensing (``$g\kappa^{\rm g}$'', red) for the same redshift bins of lenses and sources. Note, the former has been divided by a factor of $5/\ln(10)$, hence the two curves agree. The cross power spectrum for the galaxy and CMB lensing for the same lens galaxies is also shown as the green curve. The error bands including both sample variance and shot noise are forecasted according to Eq. (\ref{eq:cov}), with appropriate adaptions for $g\kappa^{\rm g}$ and $g\kappa^{\rm CMB}$ as discussed in Sec.~\ref{sec:theory}. For the auto power spectrum of CMB lensing and the reconstructed noises for the Stage IV CMB experiment as considered here, we refer the readers to~\cite{Seljak+18}.

For CMB lensing, we use the public available package $quicklens$~\cite{quicklens} to do lensing reconstruction from the T, E and B modes of CMB, and we cut off the multipoles at $\ell=3000$, due to the difficulty of cleaning temperature foregrounds at $\ell>3000$ for ground-based CMB experiments. For $g\kappa^{\rm SN}$ and $g\kappa^{\rm g}$, since we marginalize over parameters directly describing the galaxy-matter power spectrum, there is no need to worry about nonlinear effects and baryonic effects on small scales, we use information on angular scales up to $\ell\sim 10^{5}$ \cite{Zhang++05}. Of course, not much information can be obtained from modes with high enough multipoles due to significant noise.

If we define the total $S/N$ square for the galaxy-lensing cross-correlation as
\begin{equation}
\left(\frac{S}{N}\right)^2\equiv\sum_{\ell}\sum_{\substack{(LS),\\ (L'S')}} C_{\ell}^{LS} ({\rm Cov}_{\ell})^{-1}C_{\ell}^{L'S'},
\end{equation}
we find for the LSST, the galaxy-supernovae cross-correlation has $S/N=103.8$, while the galaxy-galaxy lensing has $S/N=2291.2$, about $22$ times larger due to its wider redshift coverage and smaller shot noise (see Table~\ref{tab:S/N}). While for the LSST and Stage IV CMB experiment, the galaxy-CMB lensing cross-correlation has $S/N=1842.3$.

\begin{table}
\caption{\label{tab:S/N} Total $S/N$ for detecting the galaxy-lensing cross-correlation from the LSST and Stage IV CMB experiments.}
\begin{center}
\begin{tabular}{l| c}
\hline\hline
 & S/N \\
\hline\hline
$g\kappa^{\rm SN}$        &103.8   \\
$g\kappa^{\rm g}$               &2291.2 \\
$g\kappa^{\rm CMB}$    &1842.3   \\
\hline
\end{tabular}
\end{center}
\end{table}

Throughout this paper, we choose our fiducial cosmological model to be the flat $\Lambda$CDM model with $\Omega_{\Lambda}=0.689$~\cite{Planck18}; when calculating $P_{\rm mm}$, we further choose $\{\Omega_b h^2, \Omega_m h^2, \sigma_8, n_s \}$=$\{0.022,0.142,0.81,0.967\}$~\cite{Planck18}, and adopt the Smith et al prescription~\cite{Smith+03} to account for the effects of non-linear evolution; when calculating $P_{gg}$ and $P_{gm}$, we assume a simple linear bias of $b=1$.

Degeneracy between dark energy parameters and curvature is expected, since they both affect the Universe's expansion rate, hence the comoving radial distance, though curvature has an additional effect on the comoving angular diameter distance. Therefore, the constraints on curvature will depend on specific assumptions about dark energy, characterized by assumptions for its equation of state $w$ which determines the evolution of its energy density. 

For comparison, we forecast the constraints on curvature for a wide range of choices of $w$. We first consider the following common choices: (1) $w$ is fixed to be (-1), i.e., dark energy is the cosmological constant; (2) $w$ is a constant whose value needs to be determined; (3) $w$ varies with time as $w=w_0+w_a(1-a)$, with the values of $w_0$, $w_a$ to be determined. From the first to the third choice, the degrees of freedom in $w$ is increasing. Next, we consider a fourth more general parametrization where $w$ is a binned function parametrized by its values in $N$ equal-sized $a$ bins from $a=0$ to $a=1$, which we denote as $w_i$, with $i=1,...,N$. Each $w_i$ is allowed to vary independently. 
This parametrization potentially can enclose all possible models for $w$, hence the constraints on curvature obtained with this choice after marginalizing over all dark energy parameters can be considered as independent of models of dark energy. In the following, 
we will study the constraints on curvature with $N=$10, 20, and 50, and the results when $N=50$ will be our most dark energy independent constraints on curvature.

The constraints on curvature with the above assumptions for dark energy forecasted for the LSST and Stage IV CMB experiments are shown in Table~\ref{tab:Constraints}. In different columns, we show the constraints from different probes, with the second column showing the constraints from the supernovae Hubble diagram alone (``SN''), and the third and fourth columns giving the constraints from the combinations of ``SN'' with ``$g\kappa^{\rm SN}$'' and ``$g\kappa^{\rm g}$'' respectively. The last column gives the constraints from the combinations of ``SN'' with ``$g\kappa^{\rm g}$'' and ``$g\kappa^{\rm CMB}$''. We do not show the constraints from either ``$g\kappa^{\rm SN}$'', ``$g\kappa^{\rm g}$'' or ``$g\kappa^{\rm CMB}$'' alone, because they are less interesting due to severe parameter degeneracy \cite{Bernstein06}. In different sections, we show the constraints with different assumptions for dark energy. From the first to the third sections, we show in turn the constraints with our first three choices of $w$, while from the fourth to the last, we give the constraints for the fourth choice with $N=10,\ 20,\ 50$ respectively. Note, from the top to the bottom sections, the uncertainty in our knowledge about $w$ is increasing.

\begin{table}
\caption{\label{tab:Constraints}Parameter constraints forecasted for the LSST and Stage IV CMB experiments. From left to right, the columns are constraints from the supernovae Hubble diagram (``SN''), its combination with galaxy-supernovae cross-correlation (``SN + $g\kappa^{\rm SN}$''), and combination with galaxy-galaxy lensing (``SN + $g\kappa^{\rm g}$''), and combination with galaxy-galaxy lensing and galaxy-CMB lensing(``SN + $g\kappa^{\rm g}$ + $g\kappa^{\rm CMB}$''). Note, our constraints from ``$g\kappa^{\rm SN}$'' , ``$g\kappa^{\rm g}$''  and "$g\kappa^{\rm CMB}$"are pure geometrical. 
From top to bottom, the constraints in different sections are based on dark energy parametrizations with more degrees of freedom in its equation of state $w$: the first three sections are for (1) $w=-1$, (2) $w=$const, (3) $w=w_0+w_a(1-a)$; while the last three sections assume $w$ is a binned function parametrized by its values in $N$ bins from $a=0$ to $a=1$, with $N=10,\ 20,\ 50$ respectively.}
\begin{center}
\begin{tabular}{c| c| c| c| c}
\hline\hline
parameters &SN &SN+$g\kappa^{\rm SN}$ &SN+$g\kappa^{\rm g}$ &SN+$g\kappa^{\rm g}$+$g\kappa^{\rm CMB}$ \\
\hline\hline
$\Omega_K$       &0.00815 &0.00815  &0.00659   &0.00166 \\
$\Omega_{\Lambda}$ &0.00568 &0.00568  &0.00463  &0.00142 \\
\hline
$\Omega_K$       &0.0701  &0.0698    &0.00874 &0.00200  \\
$\Omega_{\Lambda}$ &0.0720  &0.0717    &0.00882  &0.00396 \\
$w$            &0.0825  &0.0822    &0.0123 &0.0101  \\
\hline
$\Omega_K$       &0.408  &0.356    &0.0151  &0.00214 \\
$\Omega_{\Lambda}$ &0.517  &0.451    &0.00886  &0.00792 \\
$w_0$            &0.515  &0.449    &0.0156  &0.0101 \\
$w_a$            &1.28  &1.12    &0.173  &0.107 \\
\hline
$\Omega_K$       &1.16 &0.614 &0.0359 &0.0343  \\
$\Omega_{\Lambda}$ &1.30 &0.702 &0.0735  &0.0709  \\
\hline
$\Omega_K$       &8.22  &0.722 &0.0407  &0.0389  \\
$\Omega_{\Lambda}$ &9.19  &1.32 &0.175  &0.170 \\
\hline
$\Omega_K$       &9.72  &0.723 &0.0417  &0.0400  \\
$\Omega_{\Lambda}$ &8.68  &2.35  &0.391  &0.384  \\
\hline
\end{tabular}
\end{center}
\end{table}

By comparing the constraints in different sections for each probe, we find that when there are more degrees of freedom in dark energy's equation of state, the constraints on curvature get weaker, consistent with one's expectation. Our strongest constraints are hence obtained when $w=-1$, and the weakest ones obtained when $w$ is a binned function with 50 bins in $a$. It can also be easily seen that the constraints from the combined probes of either ``SN + $g\kappa^{\rm SN}$'' or ``SN + $g\kappa^{\rm g}$'' are better than ``SN'' alone, due to the important complimentary information provided by the galaxy-lensing cross-correlation which breaks the degeneracy between curvature and dark energy parameters. 

It is interesting to find that the improvements in the curvature constraints by adding in ``$g\kappa^{\rm SN}$'' or ``$g\kappa^{\rm g}$'' get more significant when dark energy has more degrees of freedom, with the improvements being minimal when dark energy is known to be the cosmological constant and maximal when $w$ is parametrized by 50 $w_i$s. This highlights the importance of galaxy-lensing cross-correlation in obtaining dark energy independent constraints on curvature. Moreover, the constraints from the combination of ``SN + $g\kappa^{\rm g}$'' are better than from ``SN + $g\kappa^{\rm SN}$''. For $N=50$, the improvement factor is 233 by adding in ``$g\kappa^{\rm g}$'' to ``SN'', while 13 when adding in ``$g\kappa^{\rm SN}$''. This is expected as compared to ``$g\kappa^{\rm SN}$'', ``$g\kappa^{\rm g}$'' has a wider redshift coverage and lower shot noise. To tighten the constraints more, we add in ``$g\kappa^{\rm CMB}$" to ``SN + $g\kappa^{\rm g}$''  in the last column, we do not add ``$g\kappa^{\rm SN}$'' here to avoid the strong correlations between ``$g\kappa^{\rm SN}$'' and ``$g\kappa^{\rm g}$'' . We notice, the improvements by adding in ``$g\kappa^{\rm CMB}$" to ``SN + $g\kappa^{\rm g}$'' get milder when dark energy has more degrees of freedom, ranging from 4 when dark energy is known to be the cosmological constant to 1.04 when $N=50$.

Finally, we find that though the curvature constraints keep getting weaker when $w$ is allowed to have more degrees of freedom, the constraints from the combined probes of ``SN + $g\kappa^{\rm g}$ + $g\kappa^{\rm CMB}$'' do not degrade much when we take $w$ to be a binned function and increase the number of bins from 10 to 20 to 50. As discussed in \cite{Witzemann+}, a limited number of bins that is equally-spaced in scale factor is enough for parameter constraints to converge. Therefore, we hereafter quote the constraints with $w$ a binned function with 50 bins in $a$ as our final ``dark energy independent'' constraints.

To summarize, we obtain dark energy independent constraints on $\Omega_K$ of $0.723$ from the combination of ``SN + $g\kappa^{\rm SN}$'' , $0.0417$ from ``SN + $g\kappa^{\rm g}$'' for the LSST, and $0.04$ from ``SN + $g\kappa^{\rm g}$ + $g\kappa^{\rm CMB}$'' for the LSST and Stage IV CMB experiments. We find the galaxy-lensing cross-correlation plays a significant role in obtaining these results. It improves the curvature constraints by breaking the degeneracy between curvature and the dark energy parameters from ``SN'' alone. If we know dark energy to be the cosmological constant, we are able to get much tighter constraints of $0.00815$ for ``SN + $g\kappa^{\rm SN}$'', $0.00659$ for ``SN + $g\kappa^{\rm g}$'', and $0.00166$ for ``SN + $g\kappa^{\rm g}$ + $g\kappa^{\rm CMB}$''. We note that in obtaining these results, we do not apply any priors. These results do not reach the desired accuracy level of $10^{-4}$, so below we add other geometrical probes to improve the constraints on curvature further.

\subsection{Adding in BAO and CMB}
\label{sec:BAO}

In this section, we are interested in tightening the constraints on curvature further by combining with other geometrical probes. Specifically, we include the probes utilizing the standard ruler of sound horizon at recombination $s$. Measurements of the CMB anisotropies can probe the angular size extended by the sound horizon at recombination $\theta_*$, hence the angular diameter distance to recombination $r_*(=s/\theta_*)$. Late-time BAO measurements can also probe the sound horizon\footnote{Strictly speaking, late-time BAO measurements probe the sound horizon at the end of the baryon drag epoch. We here neglect the small difference following \cite{Weinberg+12}.} through its imprints on matter distribution. Its extensions in the transverse direction ($\delta\theta=s/r$) and the line-of-sight direction ($\delta z =sH$) probe the late-time angular diameter distance $r$ and Hubble expansion rate $H$ respectively. In this section, we include these two probes in our forecast.

For the CMB constraints, we simply incorporate the Planck measurement for $\theta_*$, which is at an accuracy level of $\sim 0.03\%$, and is very stable to changes in the assumed cosmology~\cite{Planck18}. We calculate the sound horizon and redshift of recombination according to the fitting formulas given in~\cite{Hu05}. These two quantities are determined by the baryon and matter densities in the Universe, which themselves are well measured from morphology of the CMB anisotropy power spectrum. Therefore, we adopt the Planck constraints on $\Omega_m h^2$ and $\Omega_b h^2$ as priors for our calculation.

For the BAO constraints, we follow~\cite{Weinberg+12} and consider a Stage IV BAO experiment that maps $25\%$ of the full sky from $z=0$ to $z=3$ with errors $\sim 80\%$ larger than the linear theory sample variance errors (to account for non-negligible shot noise and non-linear degradation of the BAO signal). Such an experiments can be collectively achieved by the BAO programs that are currently ongoing or under design, such as the programs from Euclid~\cite{euclid}, WFIRST~\cite{wfirstw}, and DESI~\cite{desi}. We adopt the forecasted covariance matrix for the measured quantities of $r/s$ and $sH$ from~\cite{Weinberg+12}, and then use the Fisher matrix to derive constraints on the cosmological parameters.

The constraints we obtained from BAO and CMB  are shown in Table~\ref{tab:CombineAll}. By adding BAO and CMB to our previous calculations we get stronger constraints, also shown in Table~\ref{tab:CombineAll}. Here we only show the constraints obtained with the assumption that dark energy is the cosmological constant (top section), and with our most uncertain assumption for dark energy (i.e. $w$ is a binned function with 50 equal-sized bins from $a=0$ to $a=1$, bottom section). Note, whenever CMB is included, we apply a weak Gaussian prior with width $\Delta w_i = 10\sqrt{N}$ on all the $w_i$s \cite{Weinberg+12}.

\begin{table}
\caption{\label{tab:CombineAll} Constraints from ``BAO + CMB'', ``SN + BAO + CMB" and ``SN + $g\kappa^{\rm g}$ + $g\kappa^{\rm CMB}$ + BAO + CMB''(denoted as ``All"). The upper section assumes dark energy is the cosmological constant, while the lower section assumes $w$ is a binned function parametrized by 50 $w_i$s - its values in $50$ bins from $a=0$ to $a=1$. For each $w_i$, a prior of $\Delta w_i = 10\sqrt{N}(N=50)$  is applied.}
\begin{center}
\begin{tabular}{c| c| c| c}
\hline\hline
parameters &BAO+CMB &SN+BAO+CMB &All \\
\hline\hline
$\Omega_K$       &0.000644  &0.000514    &0.000496\\
$\Omega_{\Lambda}$ &0.00288  &0.000783     &0.000782 \\
\hline
$\Omega_K$       &0.0227  &0.00938      &0.00130 \\
$\Omega_{\Lambda}$ &0.264 &0.00939      &0.00315\\
\hline
\end{tabular}
\end{center}
\end{table}

We focus on discussions about the ``dark energy-independent constraints on curvature'', which we simply refer to as ``constraints on curvature'' unless otherwise explicitly stated in the following. From Table~\ref{tab:CombineAll}, it can be seen that our constraint on curvature from ``BAO + CMB'' is better than that from ``SN + $g\kappa^{\rm g}$ + $g\kappa^{\rm CMB}$''. Therefore, ``BAO + CMB'' is more promising in constraining curvature in a dark-energy independent way than the combination of ``SN'' and galaxy-lensing cross correlations.

At the same time, we also look at the constraint on curvature from BAO alone\footnote{Planck priors on $\Omega_mh^2$ and $\Omega_bh^2$ are still included here.}, which we find to be $0.0522$. This is slightly worse than the constraint from  ``SN + $g\kappa^{\rm g}$'' and ``SN + $g\kappa^{\rm g}$ + $g\kappa^{\rm CMB}$''. Therefore, for Stage IV dark energy surveys, the combination of ``SN'' and galaxy-lensing cross correlations provides a slightly better probe of curvature than BAO. 

To see the importance of galaxy-lensing cross correlations when they are added to the combined probe of ``SN + BAO + CMB'', we compare the 3rd and 4th columns of Table~\ref{tab:CombineAll}, and find they are only mildly helpful when dark energy is the cosmological constant, but can tighten the constraint on curvature by approximately a factor of 7 when dark energy has a general parametrization. Therefore galaxy-lensing cross-correlations play an important role in extracting information on the Universe's curvature in a dark energy independent way.

Our ultimate constraint on curvature from the combination of the five geometrical probes ``SN + $g\kappa^{\rm g}$ + $g\kappa^{\rm CMB}$ + BAO + CMB'' now reaches $0.0013$. This is much better than either ``SN + $g\kappa^{\rm g}$ + $g\kappa^{\rm CMB}$'' or ``BAO + CMB'', which give curvature constraints only at the level of $10^{-2}$, reflecting the strong complementarity of the two combined probes.

To conclude, when BAO and CMB are included, the improvements on the curvature constraints can be one order of magnitude. We find the dark energy independent constraints on curvature now can be as tight as $\sim1\times10^{-3}$, but still one order of magnitude away from the desired level of $10^{-4}$. Moreover, even if dark energy is the cosmological constant, the constraint on curvature from the combination of the five geometrical probes of ``SN + $g\kappa^{\rm g}$ + $g\kappa^{\rm CMB}$ + BAO + CMB'' is only $\sim 0.0005$, still larger than the target precision of $1\times10^{-4}$. In the following we explore how much we can gain on the curvature constraint by broadening the redshift coverage of supernovae.

\subsection{Increasing $z_{\rm max}$ of Supernovae}
\label{sec:zmax}

It is known that for a supernovae alone probe, a broader redshift coverage can give better parameter constraints provided the total number of supernovae is kept fixed~\cite{HutTur01}. This is because supernovae at different redshifts usually lead to different degeneracy directions among the parameter space, and a wider redshift coverage results in better complementarity. In this section we investigate the possibility of tightening the curvature constraints more with a wider redshift coverage of supernovae or standard candles in general.

Specifically, we keep the total number of supernovae to be fixed at $4\times10^5$, and assume their redshift distribution follows that of the LSST galaxies but cuts off at $z_{\rm max}$. In the following, we will forecast the constraints on curvature for two choices of $z_{\rm max}$: $z_{\rm max}=2$ and $z_{\rm max}=3$. Supernovae at such high redshifts may be challenging to be surveyed with ground-based telescopes like the LSST, but may be easier to observe with future space-based ones. For example, the WFIRST mission is about to find supernovae to $z_{\rm max}=1.7$, but with $N_{\rm tot}$ only $\sim 3000$~\cite{WFIRST}, far less than $4\times10^5$.  Thus, we note the forecasts we make in this section may be too optimistic for type Ia supernovae surveys currently in plan. However, they may be more realistic for other types of standard candles such as quasars and gamma-ray bursts, which can be observed to much higher redshifts \cite{King+14,Risaliti&Lusso19}.

\begin{figure}
\begin{center}
{\includegraphics[angle=0, scale=0.47]{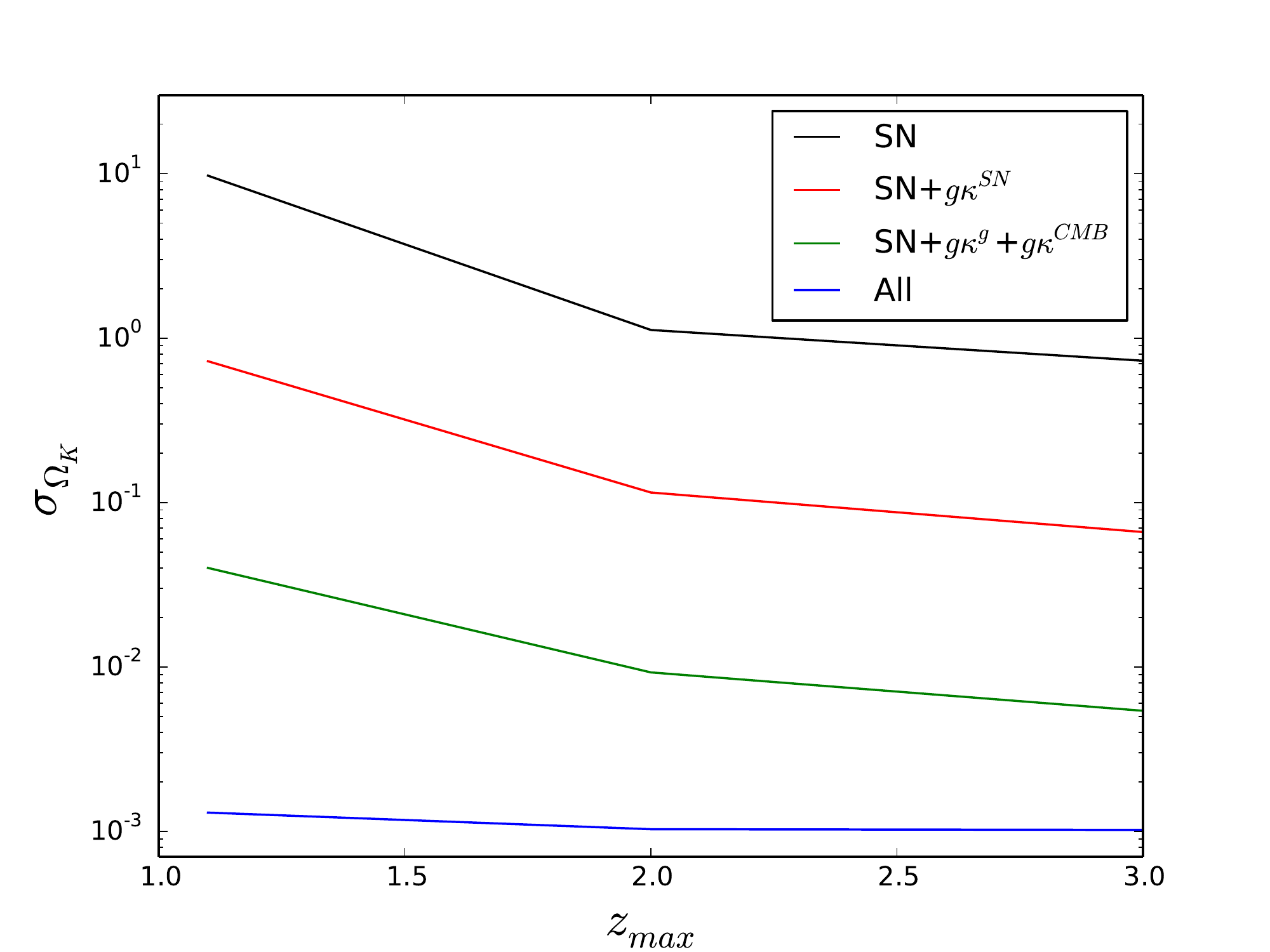}}
 \caption{\label{fig:zmaxk} Forecast 1$\sigma$ constraints on curvature as a function of the maximal redshift of supernovae for ``SN''(black) , ``SN + $g\kappa^{\rm SN}$''(red), ``SN + $g\kappa^{\rm g}$ + $g\kappa^{\rm CMB}$''(green) and ``SN + $g\kappa^{\rm g}$ + $g\kappa^{\rm CMB}$ + BAO + CMB''(blue, denoted as ``All"), respectively. The total number of supernovae is kept fixed at $4\times10^5$. The dark energy equation of state parameter $w$ is assumed to be a binned function parametrized by its values in $50$ bins from $a=0$ to $a=1$.}
\end{center}
\end{figure}

The dark energy independent constraints on curvature obtained by extending $z_{\rm max}$ from $z_{\rm max}=1.1$ to $z_{\rm max}=2$ and $z_{\rm max}=3$ are plotted in Fig.~\ref{fig:zmaxk}. It can be easily seen that the curvature constraints from either ``SN'', ``SN+$g\kappa^{\rm SN}$'', ``SN + $g\kappa^{\rm g}$ + $g\kappa^{\rm CMB}$ '', or ``SN + $g\kappa^{\rm g}$ + $g\kappa^{\rm CMB}$ + BAO + CMB'' all get better when the supernovae's redshift distribution can reach a higher $z_{\rm max}$. We also find that the improvements in the constraints by increasing $z_{\rm max}$ from $z_{\rm max}=1.1$ (Table~\ref{tab:Constraints}) to $z_{\rm max}=2$ are more significant than increasing it from $z_{\rm max}=2$ to $z_{\rm max}=3$. For the dark energy independent constraints on curvature from ``SN '', we obtain $1.12$ with $z_{\rm max}=2$ (improved by a factor of $\sim8.7$ from $z_{\rm max}=1.1$), and $0.728$ with $z_{\rm max}=3$ (improved by a factor of $\sim 1.5$ from $z_{\rm max}=2$); while for the dark energy independent constraints from ``SN + $g\kappa^{\rm g}$ + $g\kappa^{\rm CMB}$'', we get $0.00926$ with $z_{\rm max}=2$ (improved by a factor of $\sim4.3$ from $z_{\rm max}=1.1$), and $0.00541$ with $z_{\rm max}=3$ (improved by a factor of $\sim 1.7$ from $z_{\rm max}=2$). Even if we can increase the supernovae's redshift distribution only to $z_{\rm max}=2$, the efforts are rewarding enough judging from the improvements on curvature constraints from ``SN'' and its combination with galaxy-lensing cross-correlations. Again, we find as before that combining galaxy-lensing cross-correlations and ``SN'' resulting in significant improvement, even for the case with $g\kappa^{\rm SN}$, which has a relative lower $S/N$.

However, for the combination of all five probes ``SN + $g\kappa^{\rm g}$ + $g\kappa^{\rm CMB}$ + BAO + CMB'', the improvement from increasing $z_{\rm max}$ is much milder, especially when increasing from $z_{\rm max}=2$ to $z_{\rm max}=3$, reflecting the subdominant role of ``SN'' in the combined probes (probably because BAO already provides the high-$z$ information up to $z=3$), and the rapid decrease with redshift of galaxy distribution at high $z$. We find the constraints from the combination of the five probes is still $\sim 10^{-3}$. However, significant improvement on curvature constraint can be possible with standard candles whose redshift distribution has a larger fraction at high redshift, say $z>3$, which we postpone for future study.

\section{discussion}
\label{sec:discuss}

Our calculations above have adopted several simplifications. We have neglected a few systematic effects such as photometric redshift errors, shear calibration errors, galaxy intrinsic alignments etc. For photometric redshift errors, the LSST galaxies' photometry will have high enough quality to provide a rms accuracy $\sigma/(1+z)$ of 0.02 \cite{LSSTSci}. This is in general much smaller compared to our bin width of $\Delta z=0.1$. Therefore, we expect photometric redshift errors would not change at least the order of magnitude of our results. For shear calibration errors, recent analysis by \cite{Prat2019} found that if the multiplicative shear bias $m$ from LSST can be calibrated to $\sigma(m) = 0.001$, which is the requirement set in the LSST science book \cite{LSST}, its effect on cosmological constraints would be negligible. We then assume LSST shear calibration will be accurate enough not to change much of the cosmological constraints we have obtained. For intrinsic alignment, which probably will weaken our constraints on curvature to some extent, the effect can be minimized if the redshift separation between the lens and source galaxies are increased to be large enough. Considering the ultimate constraints on curvature we obtain is at the level of $10^{-3}$, one order of magnitude larger than the desired level of $10^{-4}$, we do not analyze the effect of intrinsic alignment together with other systematic effects (including those we have not mentioned in the above, such as the narrow lens bin approximation, lensing dilution and galaxy lensing boost factors) quantitatively here, which will not change our primary finding.

In this work, we obtain our dark energy-independent constraints on curvature by assuming the equation of state of dark energy is parametrized by its values in $50$ $a$ bins from $a=0$ to $a=1$ and marginalizing over all these parameters. Compared to previous work on this topic by \cite{Knox06} and \cite{Mortonson09}, this approach is more model-independent: the method proposed by \cite{Knox06} depends on the assumption that dark energy is completely negligible in ``matter-dominated'' regime, while ours allows dark energy to be non-negligible even at early times (early dark energy); \cite{Mortonson09} parametrizes $w$ of dark energy at low redshift ($z\lsim 1.7$) with $15$ principle components (PCs), but calculations of the PCs typically depend on the specifics of both the data set and cosmological model used to obtain the Fisher matrix from which they are derived. 

However, compared to previous work by \cite{Bernstein06}, our approach is not as model-independent. \cite{Bernstein06} probes curvature purely from the relationship between $r_{L}$, $r_{LS}$ and $r_S$. By marginalizing over the distances of $r_{L}$, $r_{S}$ which are integrals of functions of the Hubble expansion rate, the obtained constraints do not depend on any energy component of the Universe or dynamics that governs its expansion, but only on the validity of the FRW metric. Our approach is less general in the sense that we in addition assume energy-momentum conservation and the validity of the Friedmann equation if cosmic acceleration is due to dark energy, or if cosmic acceleration is due to modified gravity, its effect on the Universe's expansion can be viewed equivalently as an effective dark energy, which holds for most interesting modified gravity models (see e.g., \cite{PPF,Fang+08a}). Therefore, though in this paper we use the term of ``dark energy''-independent constraints on curvature, our constraints are actually independent on the unknown mechanism for cosmic acceleration including both dark energy and modified gravity.

Since we obtain our curvature constraints by marginalizing over the contribution of dark energy or ``effective'' dark energy to the Universe's expansion, while \cite{Bernstein06} obtain theirs by marginalizing over the distances, our constraints will be tighter than theirs. We conclude that \cite{Bernstein06} provide a pure metric probe of curvature which does not depend on how the Universe expands, while we probe the curvature in a way that is independent on how (``effective'') dark energy affects the Universe's expansion.

There are also many works in the literature that utilize measurements of the angular diameter distances and Hubble expansion rates to obtain model-independent constraints on curvature, e.g. \cite{Clarkson+07,Takada&Dore15,Cai+16,Yu&Wang16,Wei&Wu17,Wang+20}. These works typically need to estimate derivatives of the angular diameter distances or to reconstruct the Hubble expansion history using some model-independent smoothing techniques such as the Gaussian process. Thus, it may be hard to achieve an accuracy as tight as $\sigma_{\Omega_K}\sim10^{-4}$ robustly using these methods.

In this work, we have focused on constraining curvature using probes of the Universe's geometry. One can surely add in probes of the Universe's growth of structure to tighten the constraints. Actually, \cite{Mortonson09} has studied model independent constraints on curvature from combining geometry probes with growth probes through measuring the abundance of X-ray clusters. However, their work is done within the ``smooth'' dark energy paradigm \cite{Fang+08b} and assumes the validity of general relativity, hence it does not apply to dark energy with nontrivial clustering properties or modified gravity. Also, probes of structure growth are usually subject to systematics from theoretical modeling of baryonic physics and growth of structure on nonlinear scales. Future work on using probes of structure growth to constrain curvature in a dark energy-independent way should take into account of all these problems, which may be challenging.

\section{summary}
\label{sec:summary}

Accurate constraints on curvature provide a powerful probe of inflation. However, current accurate constraints on curvature are almost all derived upon simple assumptions of dark energy such as assuming it is the cosmological constant. Considering the large uncertainties in our theoretical understanding about dark energy, constraints with these assumptions may lead to unreliable conclusions when they are used to test inflation models. Hence, for a robust test of inflation models, it is important to obtain constraints on curvature that are independent on uncertainties in our knowledge about dark energy. In this paper, we have investigated such constraints on curvature from the geometrical probe constructed from galaxy-lensing cross-correlations and its combination with other common geometrical probes.

 We study the galaxy-magnification, galaxy-shear, and galaxy-CMB lensing cross-correlations, with magnification measured from the type Ia supernovae's brightnesses. We find for the Stage IV dark energy survey of LSST and the Stage IV CMB survey, the galaxy-magnification cross-correlation (``$g\kappa^{\rm SN}$'') can be detected with signal-to-noise ratio $S/N=104$, the galaxy-shear cross-correlation (``$g\kappa^{\rm g}$'') with $S/N=2291$, and the galaxy-CMB lensing cross-correlation (``$g\kappa^{\rm CMB}$'') with  $S/N=1842$. We include the supernovae Hubble diagram (``SN'') to break parameter degeneracy, which is available with the same supernovae data set used to measure ``$g\kappa^{\rm SN}$''. We obtain dark energy independent constraints on $\Omega_K$ of $0.723$ from ``SN + $g\kappa^{\rm SN}$'', $0.0417$ from ``SN + $g\kappa^{\rm g}$'', and $0.04$ from ``SN + $g\kappa^{\rm g}$ + $g\kappa^{\rm CMB}$'' for the LSST and Stage IV CMB experiment. We find that the galaxy-lensing cross-correlation plays a significant role in tightening the curvature constraint by breaking the degeneracy between curvature and the dark energy parameters, especially when dark energy is completely unknown. 
 We find the constraint from ``SN + $g\kappa^{\rm g}$ + $g\kappa^{\rm CMB}$'' is better than that from a Stage IV BAO experiment, but not as good when BAO is combined with the Planck measurement for the acoustic scale in the CMB. Adding the galaxy-lensing cross-correlations to the combined probe of ``SN + BAO + CMB'' results in a factor of 7 improvement in the dark energy independent constraints on curvature, but much milder improvement when dark energy is known to be the cosmological constant. We obtain our ultimate constraint on $\Omega_K$ of $0.0013$ from ``SN + $g\kappa^{\rm g}$ + $g\kappa^{\rm CMB}$ + BAO + CMB''. Our analysis also shows that tighter constraints can be obtained with better knowledge about dark energy.

We investigate the possibility of tightening the curvature constraints further by increasing the redshift extension of supernovae or standard candles in general, while keeping the total number fixed at the same value. We find though the ``SN'' alone and its combination with galaxy-lensing cross correlations have significant improvements on curvature constraints, the combined probes of ``SN + $g\kappa^{\rm g}$ + $g\kappa^{\rm CMB}$ + BAO + CMB'' does not. However, improvements can still be achievable with a larger fraction of standard candles at high redshift, larger than that for the LSST galaxies which we have assumed for the supernovae in our analysis. While this is hard to realize with supernovae, it can be easier to achieve with other types of standard candles such as quasars \cite{Risaliti&Lusso19}. We plan to investigate more about this in a future paper.

\section*{Acknowledgements}

We thank Neal Dalal and Lam Hui, discussions with whom initiated this work, and thank Pengjie Zhang, Jun Zhang, Bhuvnesh Jain for useful conversations. This work is supported by the National Natural Science Foundation of China Grants No.11773024, No.11653002, No.11421303, by the China Manned Space Program through its Space Application System, by the Fundamental Research Funds for Central Universities, and by the CAS Interdisciplinary Innovation Team.

\bibliography{omk}

\end{document}